\documentclass[conference]{IEEEtran}
\usepackage{float}
\usepackage{booktabs}
\usepackage{amsmath}
\usepackage{amsthm}
\usepackage{amssymb}
\usepackage{graphicx}
\usepackage{bm}

\makeatletter

\DeclareTextSymbolDefault{\textquotedbl}{T1}
\providecommand{\tabularnewline}{\\}
\floatstyle{ruled}
\newfloat{algorithm}{tbp}{loa}
\providecommand{\algorithmname}{Algorithm}
\floatname{algorithm}{\protect\algorithmname}

\theoremstyle{plain}
\newtheorem{thm}{\protect\theoremname}
\theoremstyle{plain}
\newtheorem{lem}[thm]{\protect\lemmaname}

\IEEEoverridecommandlockouts
\usepackage{mathrsfs}
\usepackage{amsfonts}
\usepackage{ifpdf}
\usepackage{cite}

\usepackage{array}
\usepackage{mdwmath}
\usepackage{mdwtab}
\usepackage{eqparbox}
\usepackage{subfig}

\usepackage{algorithmic}
\usepackage{algorithm}
\usepackage{epsfig}
\usepackage{times}
\usepackage{threeparttable}%add3
\usepackage{multirow}%add4
\usepackage{xcolor}
\usepackage{amsfonts}\usepackage{graphicx}
\usepackage{textcomp}
\usepackage{xcolor}
\usepackage[normalem]{ulem}
\usepackage{tabularx}
\allowdisplaybreaks[4]

\def\BibTeX{{\rm B\kern-.05em{\sc i\kern-.025em b}\kern-.08em
		T\kern-.1667em\lower.7ex\hbox{E}\kern-.125emX}}

\usepackage{amsthm}
\@ifundefined{showcaptionsetup}{}{%
 \PassOptionsToPackage{caption=false}{subfig}}
\usepackage{subfig}
\makeatother

\providecommand{\lemmaname}{Lemma}
\providecommand{\theoremname}{Theorem}

\begin{document}
\title{Task-oriented Age of Information for Remote Inference with Hybrid Language Models\\ % Small and Large Language Models }
%\thanks{This work was supported in part by the National Key Research and Development Program of China under Grant 2024YFE0200300, in part by the National Natural Science Foundation of China under Grants 62271513 and 62271413, in part by the Research Fund under the Shaanxi Province Innovation Capability Support Program under Grant 2023KJXX-010, and in part by the Talents Special Foundation of Northwest A\&F University under Grant Z1090324128.}
}
\author{\IEEEauthorblockN{{Shuying Gan}\textsuperscript{1}, {Xijun Wang}\textsuperscript{1}\textsuperscript{*},
{Chenyuan Feng}\textsuperscript{2}\textsuperscript{*}, {Chao Xu}\textsuperscript{3},
{Howard H. Yang}\textsuperscript{4}, {Xiang Chen}\textsuperscript{1},
and {Tony Q. S. Quek}\textsuperscript{5}\\
} \IEEEauthorblockA{\textsuperscript{1}School of Electronics and Information Engineering,
Sun Yat-sen University, Guangzhou, China\\
 \textsuperscript{2}Department of Communication Systems, EURECOM,
Sophia Antipolis, France\\
 \textsuperscript{3}School of Information Engineering, Northwest
A\&F University, Yangling, China\\
 \textsuperscript{4}ZJU-UIUC Institute, Zhejiang University, Haining,
China\\
 \textsuperscript{5}Information System and Technology Design Pillar,
Singapore University of Technology and Design, Singapore\\
 Email: ganshy7@mail2.sysu.edu.cn, wangxijun@mail.sysu.edu.cn, Chenyuan.Feng@eurecom.fr,
\\
cxu@nwafu.edu.cn, haoyang@intl.zju.edu.cn, chenxiang@mail.sysu.edu.cn,
tonyquek@sutd.edu.sg}}

\maketitle

\begin{abstract}
Large Language Models (LLMs) have revolutionized the field of artificial intelligence (AI) through their advanced reasoning capabilities, but their extensive parameter sets introduce significant inference latency, posing a challenge to ensure the timeliness of inference results. While Small Language Models (SLMs) offer faster inference speeds with fewer parameters, they often compromise accuracy on complex tasks. This study proposes a novel remote inference system comprising a user, a sensor, and an edge server that integrates both model types alongside a decision maker. The system dynamically determines the resolution of images transmitted by the sensor and routes inference tasks to either an SLM or LLM to optimize performance.
The key objective is to minimize the Task-oriented Age of Information (TAoI) by jointly considering the accuracy and timeliness of the inference task. Due to the non-uniform transmission time and inference time, we formulate this problem as a Semi-Markov Decision Process (SMDP). By converting the SMDP to an equivalent Markov decision process, we prove that the optimal control policy follows a threshold-based structure. We further develop a relative policy iteration algorithm leveraging this threshold property. Simulation results demonstrate that our proposed optimal policy significantly outperforms baseline approaches in managing the accuracy-timeliness trade-off. 
\end{abstract}

\begin{IEEEkeywords}
Remote inference, task-oriented age of information, semi-Markov decision process, small language models, large language models
\end{IEEEkeywords}

\section{Introduction}
It is prevalent to provide artificial intelligence (AI) services in remote inference systems using status update data collected from sensors, such as in applications like intelligent transportation, industrial automation, and personal assistance \cite{Chen2024}. Within these frameworks, the execution of inference tasks relies on the transmission of data to pre-trained neural networks, where both the precision and timeliness of inference are paramount for maintaining the quality of service. Large Language Models (LLMs), celebrated for their extensive comprehension and reasoning skills, have become prominent AI services for ensuring the accuracy of inferences \cite{GSS2023,Lan2024}. However, the pursuit of enhanced accuracy in LLMs has resulted in models with an enormous parameter count, such as GPT-4 and LLaMA-405B, leading to a notable increase in inference latency. In contrast, Small Language Models (SLMs),  with reduced parameter sets, enable more rapid inference but may compromise accuracy, especially when dealing with intricate tasks \cite{Eldan2023}, as seen with models like LLaMA-7B and ALBERT. Consequently, the effective orchestration of SLMs and LLMs within remote inference systems to achieve a balance between inference accuracy and timeliness presents a challenge that merits exploration.

Previous studies have explored methods for achieving efficient and timely inference in remote inference systems. In \cite{Shisher2023}, the authors showed that inference error is not necessarily a linear function of the age of information (AoI) nor a non-increasing function of the feature length. They jointly optimized feature length selection and transmission scheduling to minimize the average inference error. Building on this work, the authors of \cite{Shisher2024} proposed a selection-from-buffer model for feature selection to reduce inference error. 
%This model retains the most recent $B$ features in the source buffer, allowing the source to transmit any of these $B$ features. 
In \cite{Ari2024}, the focus was on minimizing remote inference error for a dynamically changing objective at the receiver. The concept of task-oriented age of information (TAoI) was introduced in \cite{Gan2024} to quantify the timeliness of the inference tasks in a remote inference system with pre-discrimination at the transmitter. A limitation of the remote inference systems in the above works is that they were restricted to scenarios with a single network at the receiver, ignoring the impact of network architecture on inference performance. The authors of \cite{Zhao2023} proposed an online optimization framework for multi-user and multi-DNN inference services. This framework aimed to strike a balance between inference precision, latency, and resource expenditure by jointly optimizing DNN model selection and resource allocation. While \cite{Zhao2023} recognized the importance of accounting for multiple neural networks at the receiver, solely focusing on minimizing inference latency is insufficient to guarantee the timeliness of remote inference systems. %The comprehensive optimization of both network structure and latency is crucial for enhancing the overall performance of remote inference systems.

Motivated by these limitations, we investigate a remote inference system with hybrid SLM and LLM. In particular, the system consists of a user, a sensor, and an edge server equipped with a decision maker, an SLM, and an LLM. 
Given that different image resolutions and model sizes result in varying transmission and inference latencies as well as accuracies, the decision maker controls the resolution of the image transmitted by the sensor and decides whether to forward it to the SLM or LLM for inference. To strike a balance between timeliness and accuracy, we employ TAoI as the performance metric, which is reduced upon successful inference and accumulates otherwise, and aim to develop an optimal control strategy that minimizes the TAoI. By modeling this dynamic control problem as a finite Semi-Markov Decision Process (SMDP) and then converting it into a Markov Decision Process (MDP) with uniform time steps, we prove that the optimal policy adheres to a threshold-based structure.
Furthermore, we propose a Relative Policy Iteration (RPI) algorithm that leverages this threshold-based approach to yield the optimal control policy. Finally, simulation results verify that the proposed policy outperforms baseline strategies in terms of TAoI minimization.

%The rest of this paper is organized as follows. Section II introduces the system model. In Section III, we present the SMDP formulation and design an RVI algorithm based on the threshold structure. Section IV discusses the simulation results, followed by the conclusion in Section V.

\section{System Model}\label{Sec:Section 2} 

\begin{figure}[!t]
\centering\includegraphics[width=0.48\textwidth]{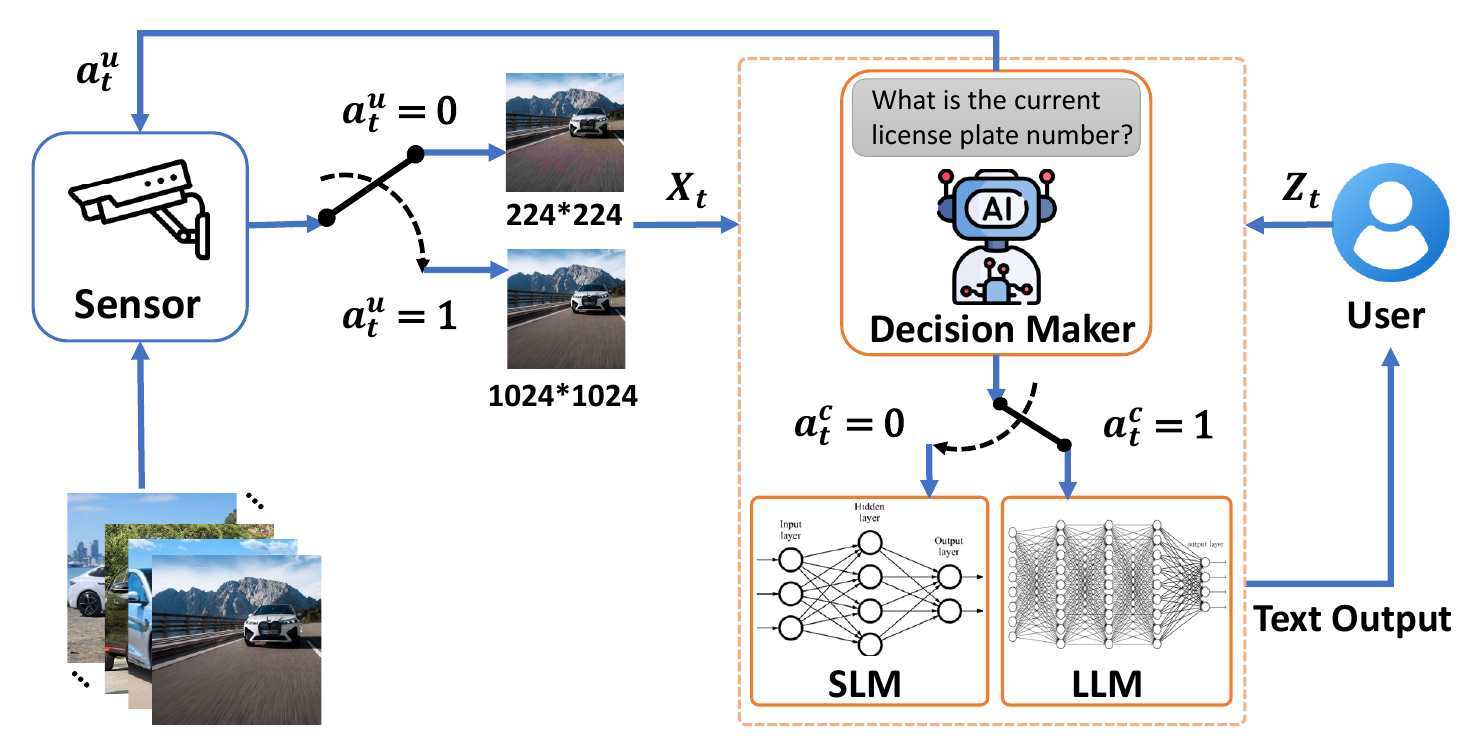}
\vspace{-0.4em}
\caption{An illustration of the remote inference system with hybrid SLM and LLM.}
\vspace{-1em}
\label{Fig:System_Model}
\end{figure}

As shown in Fig. \ref{Fig:System_Model}, we consider a remote inference system consisting of a user, a sensor, and an edge server. The sensor captures real-time scenes and generates images at resolutions determined by the edge server. These images are then transmitted to the edge server, which houses three key components: a decision maker, a SLM, and a LLM.
The system balances precision and timeliness in executing inference tasks, such as responding to queries like \textquotedbl What is the current license plate number?\textquotedbl . High-resolution images, while providing more detailed clarity beneficial for complex inference tasks, come with increased transmission latency.  For instance, when a vehicle is at a distance from the sensor, a high-resolution image is crucial for accurately identifying the license plate, despite the longer transmission time. In contrast, when a vehicle is in close proximity, a low-resolution image suffices for the inference task and results in faster transmission.  Upon receiving images, the decision maker decides whether the current inference task should be handled by the SLM or the LLM. 
The SLM offers faster processing with moderate accuracy, while the LLM provides higher accuracy at the cost of increased processing time. The selected model generates text output for the user, who provides satisfaction feedback to the decision maker. This feedback loop enables continuous refinement of the inference process.

We consider that the system is time-slotted, where each time slot lasts for a duration of $\tau$ seconds. A decision epoch of the decision maker is denoted as a discrete time step $t$, and each time step contains multiple time slots.  At the beginning of each time step, the decision maker selects the resolution of the image to be transmitted and decides on the language model to be used. Let $a_{t}^{u}\in\left\{ 0,1\right\}$ denote the resolution decision, with $a_{t}^{u}=0$ indicating that the sensor is instructed to transmit a low-resolution image, and $a_{t}^{u}=1$ indicating that a high-resolution image is to be sent. Based on the resolution decision $a_{t}^{u}$, the sensor captures a real-time scene and generates an image $X_{t}\in\mathcal{X}$ at the specified resolution,  which is then transmitted to the edge server. We assume that the transmission process is reliable, with the transmission latency for a low-resolution image being $T_{1}^{u}$ and for a high-resolution image being $T_{2}^{u}$. Note that $T_{1}^{u}$ is always less than $T_{2}^{u}$.
Let \(a_{t}^{c}\) denote the inference decision, where \(a_{t}^{c} = 0\) signifies inference by the SLM, and \(a_{t}^{c} = 1\) signifies inference by the LLM. 
When the edge server receives the transmitted image $X_{t}$, the decision maker sends the image and its corresponding query to the SLM or the LLM for inference according to \(a_{t}^{c}\).  Let $T_{1}^{c}$ and $T_{2}^{c}$ denote the inference latencies for the SLM and LLM, respectively, with  \(T_{1}^{c} < T_{2}^{c}\).   Then, the control action vector of the decision maker at time step \(t\) is denoted by \(\bm{a}_{t} \triangleq (a_{t}^{u}, a_{t}^{c}) \in \mathcal{A} \triangleq \{(0,0), (0,1), (1,0), (1,1)\}\), where \(\mathcal{A}\) represents the set of all possible actions.  Note that the duration of a time step is not constant. Specifically, let \( L(\bm{a}_{t}) \) represent the number of time slots within time step \( t \) when action \( \bm{a}_{t} \) is executed. This can be formulated as:
\begin{align}
L(\bm{a}_{t})=\begin{cases}
T_{1}^{u}+T_{1}^{c}, & \text{ if }\bm{a}_{t}\text{= (0,0)};\\
T_{2}^{u}+T_{1}^{c}, & \text{ if }\bm{a}_{t}\text{= (1,0)};\\
T_{1}^{u}+T_{2}^{c}, & \text{ if }\bm{a}_{t}\text{= (0,1)};\\
T_{2}^{u}+T_{2}^{c}, & \text{ if }\bm{a}_{t}\text{= (1,1)}.
\end{cases} & \label{Eq:time_step}
\end{align}

The user, upon receiving the text output of the language model, sends feedback $Z_{t}$ to the receiver, with \(Z_{t} = 1\) indicating a correct output and \(Z_{t} = 0\) indicating otherwise. We assume that the latency associated with transmitting the text output to the user is negligible. It is important to note that the inference accuracy is influenced not only by the size of the language model but also by the image resolution. Let \(p_{s}\) and \(q_{s}\) denote the probabilities of correct inference when a low-resolution and high-resolution image, respectively, are sent to the SLM for inference, i.e., 
\vspace{-0.4em}
\begin{align}
 & p_{s}\triangleq\mathrm{Pr}(Z_{t}=1|\mathbf{a}_{t}\text{= (0,0)}),\forall t,\label{Eq:low-SLM}\\
 & q_{s}\triangleq\mathrm{Pr}(Z_{t}=1|\mathbf{a}_{t}\text{= (1,0)}),\forall t.\label{Eq:high-SLM}
\end{align}
Similarly, $p_{l}$ and $q_{l}$ are defined as the probability of correct inference when a low-resolution and high-resolution image, respectively, are transmitted to the LLM for inference, i.e.,
\vspace{-0.4em}
\begin{align}
 & p_{l}\triangleq\mathrm{Pr}(Z_{t}=1|\mathbf{a}_{t}\text{= (0,1)}),\forall t,\label{Eq:low-LLM}\\
 & q_{l}\triangleq\mathrm{Pr}(Z_{t}=1|\mathbf{a}_{t}\text{= (1,1)}),\forall t.\label{Eq:high-LLM}
\end{align}

AoI serves as a prevalent metric for quantifying the freshness of data as perceived by the receiver \cite{Yates2021}. However, it does not capture the utility of the information content with respect to the specific task. To bridge this gap, our remote inference system employs TAoI to measure the accuracy and timeliness of the inference task \cite{Gan2024}. Specifically, TAoI only decreases upon the successful completion of an inference task; in other cases, it increases. Let $U_{t}$ denote the time step at which the most up-to-date correct text output received by the user was generated. The TAoI at the $i$-th time slot of the time step $t$ is defined as $ \Delta_{t,i}=\sum_{n=U_{t}}^{t-1}L(\mathbf{a}_{n})+i-1$, where the first term represents the total number of time slots in the previous time steps since $U_{t}$. For ease of explanation,  we represent the TAoI at the beginning of time step $t$ as $\Delta_{t}$. That is, $\Delta_{t}=\Delta_{t,1}=\sum_{n=U_{t}}^{t-1}L(\mathbf{a}_{n})$. We introduce $\hat{\Delta}$ as the upper limit of the TAoI, which is assumed to be finite but can be arbitrarily large. Upon successful completion of the inference task  (i.e., $Z_{t}=1$), TAoI is reduced to its corresponding total latency. For instance, if a low-resolution image is transmitted, the SLM is selected for inference, and the text output is correct  (i.e., $\mathbf{a}_{t}= (0,0)$ and $Z_{t}=0$), then the TAoI resets to $T_{1}^{u}+T_{1}^{c}$. Conversely, if the inference task fails (i.e., $\mathbf{a}_{t}= (0,0)$ and $Z_{t}=0$ ),  TAoI increases by  $T_{1}^{u}+T_{1}^{c}$. Therefore, the evolution of TAoI can be illustrated as follows:
\vspace{-0.2em}
\begin{align}
\Delta_{t+1}=\begin{cases}
T_{1}^{u}+T_{1}^{c}, & \mathbf{a}_{t}\text{= (0,0)}\;\text{\&}\;Z_{t}=1;\\
T_{2}^{u}+T_{1}^{c}, & \mathbf{a}_{t}\text{= (1,0)}\;\text{\&}\;Z_{t}=1;\\
T_{1}^{u}+T_{2}^{c}, & \mathbf{a}_{t}\text{= (0,1)}\;\text{\&}\;Z_{t}=1;\\
T_{2}^{u}+T_{2}^{c}, & \mathbf{a}_{t}\text{= (1,1)}\;\text{\&}\;Z_{t}=1;\\
\min\{\Delta_{t}+T_{1}^{u}+T_{1}^{c},\hat{\Delta}\}, & \mathbf{a}_{t}\text{= (0,0)}\;\text{\&}\;Z_{t}=0;\\
\min\{\Delta_{t}+T_{2}^{u}+T_{1}^{c},\hat{\Delta}\}, & \mathbf{a}_{t}\text{= (1,0)}\;\text{\&}\;Z_{t}=0;\\
\min\{\Delta_{t}+T_{1}^{u}+T_{2}^{c},\hat{\Delta}\}, & \mathbf{a}_{t}\text{= (0,1)}\;\text{\&}\;Z_{t}=0;\\
\min\{\Delta_{t}+T_{2}^{u}+T_{2}^{c},\hat{\Delta}\}, & \mathbf{a}_{t}\text{= (1,1)}\;\text{\&}\;Z_{t}=0.
\end{cases}\label{Eq:AoI}
\end{align}
\vspace{-0.4em}

In this study, our objective is to develop a control policy $\pi=\left(\mathbf{a}_{1},\mathbf{a}_{2},\cdots\right)$
that minimizes the long-term average TAoI. The dynamic control problem can be formulated as follows: 
\vspace{-0.4em}
\begin{align}
\min_{\pi}\limsup_{T\rightarrow\infty}{\dfrac{\mathbb{E}\left[\sum_{t=1}^{T}\Delta_{t}\right]}{\mathbb{E}\left[\sum_{t=1}^{T}L(\mathbf{a}_{t})\right]}}.\label{Eq:Our_pro}
\end{align}

\section{SMDP Formulation and Solution}\label{Sec:Section 3} 
\subsection{SMDP Formulation}
Due to the non-uniform durations of time intervals between decision epochs, we reformulate the dynamic control problem \eqref{Eq:Our_pro} as the SMDP. Specifically, an SMDP  is composed of a tuple  $\left(\mathcal{S},\mathcal{A},t^{+},\mathrm{Pr}(\cdot,\cdot),R(\cdot,\cdot)\right)$, where each component is defined as follows:

\textbf{1) State space $\mathcal{S}$}: The state of the SMDP at time step $t$ is defined as the TAoI, denoted by $s_{t}\triangleq\Delta_{t}$. Given that the TAoI is bounded by its upper limit $\hat{\Delta}$, the state space \textbf{$\mathcal{S}$} is finite.

\textbf{2) Action space $\mathcal{A}$}: The action of the SMDP at time step $t$  comprises a resolution decision and an inference decision made by the decision maker, denoted by $\mathbf{a}_{t}\triangleq(a_{t}^{u},a_{t}^{c})$. The action space is \textbf{$\mathcal{A}$$\triangleq\left\{ (0,0),(0,1),(1,0),(1,1)\right\} $}.

\textbf{3) Decision epoch $t^{+}$}: The time interval $L(\mathbf{a}_{t})$ between two consecutive decision epochs is determined by the action $\mathbf{a}_{t}$ taken at time step $t$, as detailed in \eqref{Eq:time_step}.

\textbf{4) Transition probability $\mathrm{Pr}(\cdot,\cdot)$}: Let $\mathrm{Pr}(s_{t+1}|s_{t},\mathbf{a}_{t})$ denote the transition probability from the current state $s_{t}$ to the next state $s_{t+1}$ under action $\mathbf{a}_{t}$. According to the TAoI evolution dynamics in (\ref{Eq:AoI}), the transition probabilities are detailed in Table \ref{Ta:Tran_pro}.

\textbf{5) Cost function $R(\cdot,\cdot)$}: We define the instantaneous cost under state $s_{t}$ given action $\bm{a}_{t}$ as follows: 
\vspace{-0.4em}
\begin{align}
R(s_{t},\mathbf{a}_{t}) & =R(\Delta_{t},\mathbf{a}_{t}) =\sum_{i=1}^{L(\mathbf{a}_{t})}\Delta_{t,i}=\sum_{i=1}^{L(\mathbf{a}_{t})}\Delta_{t}+i-1\nonumber \\
 & =L(\mathbf{a}_{t})[\Delta_{t}+\dfrac{1}{2}\left(L(\mathbf{a}_{t})-1\right)].
\end{align}

\begin{table}[!t]
\global\long\def\arraystretch{1.2}%
\centering \caption{Transition probability}
\vspace{-0.2em}
\begin{tabular}{cccc}
\toprule 
$\mathrm{Pr}(s_{t+1}|s_{t},a_{t})$ & $s_{t}$ & $\mathbf{a}_{t}$ & $s_{t+1}$\tabularnewline
\midrule 
$p_{s}$ & $\Delta_{t}$ & $\left(0,0\right)$ & $T_{1}^{u}+T_{1}^{c}$\tabularnewline
$1-p_{s}$ & $\Delta_{t}$ & $\left(0,0\right)$ & $\min\{\Delta_{t}+T_{1}^{u}+T_{1}^{c},\hat{\Delta}\}$\tabularnewline
$q_{s}$ & $\Delta_{t}$ & $\left(1,0\right)$ & $T_{2}^{u}+T_{1}^{c}$\tabularnewline
$1-q_{s}$ & $\Delta_{t}$ & $\left(1,0\right)$ & $\min\{\Delta_{t}+T_{2}^{u}+T_{1}^{c},\hat{\Delta}\}$\tabularnewline
$p_{l}$ & $\Delta_{t}$ & $\left(0,1\right)$ & $T_{1}^{u}+T_{2}^{c}$\tabularnewline
$1-p_{l}$ & $\Delta_{t}$ & $\left(0,1\right)$ & $\min\{\Delta_{t}+T_{1}^{u}+T_{2}^{c},\hat{\Delta}\}$\tabularnewline
$q_{l}$ & $\Delta_{t}$ & $\left(1,1\right)$ & $T_{2}^{u}+T_{2}^{c}$\tabularnewline
$1-q_{l}$ & $\Delta_{t}$ & $\left(1,1\right)$ & $\min\{\Delta_{t}+T_{2}^{u}+T_{2}^{c},\hat{\Delta}\}$\tabularnewline
\bottomrule
\end{tabular}\label{Ta:Tran_pro} \vspace{-1em}
\end{table}

Given an initial system state $s_{1}$, the objective can be expressed as follows:
\vspace{-0.4em}
\begin{align}
\min_{\pi}\limsup_{T\rightarrow\infty}\dfrac{\mathbb{E}\left[\sum_{t=1}^{T}R(s_{t},\mathbf{a}_{t})\mid s_{1}\right]}{\mathbb{E}\left[\sum_{t=1}^{T}L(\mathbf{a}_{t})\right]} & .\label{Eq:trans_policy-1}
\end{align}
Our goal is to find a stationary deterministic optimal control policy $\pi^{*}$ that solves the long-term average TAoI minimization problem as presented in \eqref{Eq:trans_policy-1}. Before analyzing the stationary deterministic optimal policy for average TAoI, it is imperative to confirm the existence of such a policy.  According to \cite[Theorem 8.4.5]{Puterman2014}, a deterministic stationary average optimal policy exists for a finite-state finite-action average-cost MDP provided that the cost function is bounded and the MDP is unichain. Thus, we need examine the two prerequisites for the existence of a deterministic stationary policy: i) First, the cost in the MDP is bounded, as the instantaneous cost is defined by the TAoI, which is capped by an upper limit $\hat{\Delta}$; ii) Second, given that the state $\hat{\Delta}$ is accessible from every other state, our Markov chain forms a single recurrent class, signifying that the MDP is unichain. Hence, a stationary deterministic optimal policy  is confirmed to exist for this dynamic control problem.

To derive the optimal control policy, we begin by converting the SMDP into an equivalent discrete-time MDP \cite{Puterman2014}. Let $\hat{\mathcal{S}}$ and $\hat{\mathcal{A}}$ denote the state and action spaces of the transformed MDP, respectively. These spaces are identical to those of the original SMDP, that is, $\hat{\mathcal{S}}=\mathcal{S}$ and $\hat{\mathcal{A}}=\mathcal{A}$. For any state $s=\Delta\in\hat{\mathcal{S}}$
and action $\bm{a}\in\hat{\mathcal{A}}$, the cost in the MDP is
\vspace{-0.5em}
\begin{align}
\bar{R}(\Delta,\bm{a})=\frac{R(\Delta,\bm{a})}{L(\bm{a})} = \Delta+\dfrac{1}{2}(L(\bm{a})-1) & ,\label{Eq:reward}
\end{align}
and the transition probability is given by 
\begin{align}
\bar{p}(s'|s,\bm{a})=\begin{cases}
\frac{\epsilon}{L(\bm{a})}p(s'|s,\bm{a}), & s'\neq s\\
1-\frac{\epsilon}{L(\bm{a})}, & s'=s
\end{cases} & ,\label{Eq:Pro}
\end{align}
where $\epsilon$ is  selected to be within the interval $(0,\min_{\bm{a}}L(\bm{a})]$. The objective is then to find a policy $\pi\in\Pi$ that minimizes the following:
\begin{align}
\min_{\pi\in\Pi} \dfrac{1}{T}\limsup_{T\rightarrow\infty}{\mathbb{E}\left[\sum_{t=1}^{T}\bar{R}(s_{t},\bm{a}_{t})\mid s_{1}\right]} & .\label{Eq:V}
\end{align}
We focus on the set of deterministic stationary policies $\Pi$, where $\pi=\{\bm{a}_{1},\bm{a}_{2},\cdots\}\in\Pi$ such that $\bm{a}_{t_{1}}=\bm{a}_{t_{2}}$ when $s_{t_{1}}=s_{t_{2}}$ for any ${t_{1}},{t_{2}}$. For simplicity, we omit the time index in the sequel.
The optimal policy $\pi^{*}$ can be derived by solving the corresponding Bellman equation. According to \cite{Bertsekas2012}, we have:
\vspace{-0.5em}
\begin{align}
V^{*}+V(s)=\min_{\bm{a}\in\mathcal{A}}\left\{ \bar{R}(s,\bm{a})+\sum_{s'\in\mathcal{S}}\bar{p}(s'|s,\bm{a})V(s')\right\} ,\;\forall s\in\mathcal{S} & ,\label{Eq:Bellman}
\end{align}
where $V^{*}$ represents the optimal value to \eqref{Eq:trans_policy-1} for all initial states, and $V(\bm{s})$ is the value function for the discrete-time MDP. The optimal policy $\pi^{*}$ for any state $\bm{s}\in\mathcal{S}$ is given by: 
\begin{align}
\pi^{*}(s)=\arg\min_{\bm{a}\in\mathcal{A}}\left\{ \bar{R}(s,\bm{a})+\sum_{s'\in\mathcal{S}}\bar{p}(s'|s,\bm{a})V(s')\right\} ,\;\forall s\in\mathcal{S}.\label{Eq:opt_polic}
\end{align}

\subsection{Structural Analysis and Optimal Policy}
Our first step is to prove that the optimal policy exhibits a threshold-like structure. Based on this, we develop an RPI algorithm that exploits this threshold structure to find the optimal policy. To proceed, we present key properties of the value function, as shown in the following lemmas.
\begin{lem} \label{LM:Lemma1}
The value function $V(\Delta)$ is non-decreasing with $\Delta$.
\end{lem}
\begin{IEEEproof}
See Section II-A in the online materials \cite{Supplementary}.
\end{IEEEproof}
\begin{lem} \label{LM:Lemma2} 
The value function $V(\Delta)$ is concave with $\Delta$. 
\end{lem}
\begin{IEEEproof}
See Section II-B in the online materials \cite{Supplementary}.
\end{IEEEproof}
Since the value function $V(\Delta)$ is non-decreasing and concave, its slope is non-increasing and lower bounded. The lower bound of the slope of $V(\Delta)$ is given by the following lemma. Prior to that, we define an auxiliary variable $l_{min}$ as follows:
\begin{equation}
l_{min}=\min \left(\frac{T_{1}^{u}+T_{1}^{c}}{p_{s}},\frac{T_{2}^{u}+T_{1}^{c}}{q_{s}},\frac{T_{1}^{u}+T_{2}^{c}}{p_{l}},\frac{T_{2}^{u}+T_{2}^{c}}{q_{l}}\right).\label{eq:min_slope}
\end{equation}

\begin{lem} \label{LM:Lemma3} 
For any $\Delta_{1}$, $\Delta_{2}\in\mathcal{S}$ with $\Delta_{1}\leq\Delta_{2}$, we have $V(\Delta_{2})-V(\Delta_{1})\geq\dfrac{L(\hat{\mathbf{a}})}{\epsilon \hat p}(\Delta_{2}-\Delta_{1})$, where $\hat{\mathbf{a}}$ and $\hat p$ are given by
\vspace{-1em}
\begin{align}
(\hat{\mathbf{a}},\hat p)=\begin{cases}
((0,0),p_{s}), & \text{ if }l_{min}=\frac{T_{1}^{u}+T_{1}^{c}}{p_{s}};\\
((1,0),q_{s}), & \text{ if }l_{min}=\frac{T_{2}^{u}+T_{1}^{c}}{q_{s}};\\
((0,1),p_{l}), & \text{ if }l_{min}=\frac{T_{1}^{u}+T_{2}^{c}}{p_{l}};\\
((1,1),q_{l}), & \text{ if }l_{min}=\frac{T_{2}^{u}+T_{2}^{c}}{q_{l}}.
\end{cases} & \label{Eq:mid}
\end{align}
\end{lem}
\begin{IEEEproof}
See Section II-C in the online materials \cite{Supplementary}.
\end{IEEEproof}

Based on Lemmas \ref{LM:Lemma1}-\ref{LM:Lemma3}, we can derive the structure of the optimal control policy as stated in the following theorem.
\begin{thm} \label{Th:Theorem1}
For any $\Delta_{1}, \Delta_{2}\in\mathcal{S}$ with $\Delta_{1}\leq\Delta_{2}$, there exists a stationary deterministic optimal policy with a threshold-based structure, described as follows:

\noindent $\boldsymbol{\cdot}$ When $l_{min}=\frac{T_{1}^{u}+T_{1}^{c}}{p_{s}}$ and $\pi^{*}(\Delta_{1})=(0,0)$,  $\pi^{*}(\Delta_{2})=(0,0)$. 

\noindent $\boldsymbol{\cdot}$ When $l_{min}=\frac{T_{2}^{u}+T_{1}^{c}}{q_{s}}$ and $\pi^{*}(\Delta_{1})=(1,0)$,  $\pi^{*}(\Delta_{2})=(1,0)$. 

\noindent $\boldsymbol{\cdot}$ When $l_{min}=\frac{T_{1}^{u}+T_{2}^{c}}{p_{l}}$ and $\pi^{*}(\Delta_{1})=(0,1)$,  $\pi^{*}(\Delta_{2})=(0,1)$. 

\noindent $\boldsymbol{\cdot}$ When $l_{min}=\frac{T_{2}^{u}+T_{2}^{c}}{q_{l}}$ and $\pi^{*}(\Delta_{1})=(1,1)$,  $\pi^{*}(\Delta_{2})=(1,1)$.
\end{thm}
\begin{IEEEproof}
Please refer to Appendix \ref{PR:proof_Theorem1}. 
\end{IEEEproof}

\begin{figure}[!t]
\centering
\includegraphics[width=0.4\textwidth]{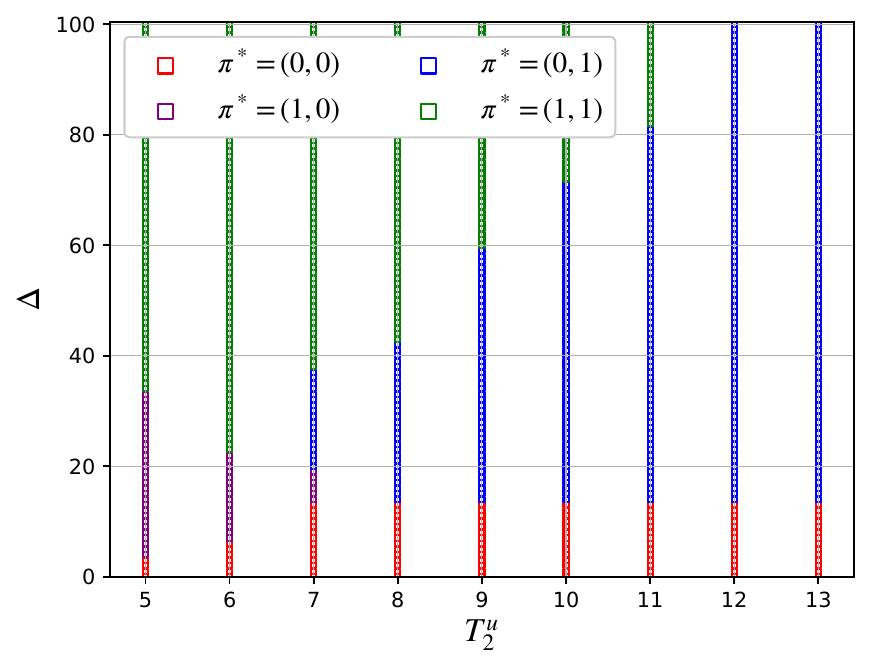}
\caption{Structure of the optimal policy for different $T_{2}^{u}$ ($T_{1}^{u}=4$, $T_{1}^{c}=3$, $T_{2}^{c}=4$, $p_{s}=0.3$, $q_{s}=0.7$, $p_{l}=0.5$, $q_{l}=0.8$).}
\label{Fig:Threshold}
\end{figure}

Theorem \ref{Th:Theorem1} shows the existence of a threshold structure within the optimal policy across four different cases. It is further verified by Fig. \ref{Fig:Threshold}, which shows that the structure of the optimal policy  corresponds to case 4 when $T_{2}^{u} \leq 11$ and to case 3 when $T_{2}^{u} > 11$. Based on this threshold structure, we propose the RPI algorithm, as outlined in Algorithm \ref{alg:algorithm2-1}. Specifically, if the condition outlined in Theorem \ref{Th:Theorem1} is satisfied, the optimal policy can be determined directly within lines 5-12 of Algorithm~\ref{alg:algorithm2-1} without the need to search through all possible actions. This significantly reduces the computational complexity of the algorithm.

\begin{algorithm}[t!]
\caption{RPI Algorithm Based on the Threshold Structure}
\label{alg:algorithm2-1} \begin{algorithmic}[1]
\STATE \textbf{Initialization:} Set $\pi_{0}^{*}(s)=0$ for all $s\in S$, select a reference state $s^{\dagger}$, and set $k=0$.
\STATE\textbf{Policy Evaluation:} Given $\pi_{k}^{*}$ and $V_{k}(s^{\dagger})$, compute $V^*_{k}$ and $V_{k}(s)$ according to $V^*_{k}+V_{k}(s)=\bar{R}(s,\pi_{k}^{*}(s))+\sum_{s^{\prime}\in\mathcal{S}}\bar{p}(s^{'}|s,\pi_{k}^{*}(s))V_{k}(s^{\prime})$.
\STATE\textbf{Policy Improvement Based on the Threshold Structure:}
Compute a new policy $\pi_{k}^{*}$ for each $s\in S$ as follows:
\FOR {$s\in S$} 
\IF{$l_{min}=\frac{T_{1}^{u}+T_{1}^{c}}{p_{s}}$ and $\pi_{k+1}^{*}(s-1)=(0,0)$} 
\STATE $\pi_{k+1}^{*}(s)=(0,0)$; 
\ELSIF{$l_{min}=\frac{T_{2}^{u}+T_{1}^{c}}{q_{s}}$ and $\pi_{k+1}^{*}(s-1)=(1,0)$} 
\STATE $\pi_{k+1}^{*}(s)=(1,0)$; 
\ELSIF{$l_{min}=\frac{T_{1}^{u}+T_{2}^{c}}{p_{l}}$ and $\pi_{k+1}^{*}(s-1)=(0,1)$} 
\STATE $\pi_{k+1}^{*}(s)=(0,1)$; 
\ELSIF{$l_{min}=\frac{T_{2}^{u}+T_{2}^{c}}{q_{l}}$ and $\pi_{k+1}^{*}(s-1)=(1,1)$} 
\STATE $\pi_{k+1}^{*}(s)=(1,1)$; 
\ELSE 
\STATE $\pi_{k+1}^{*}(s)=\arg\min_{\mathbf{a}\in\mathcal{A}}\{\bar{R}(s,\pi_{k}^{*}(s))+$
\STATE $\hspace{10 em}\sum_{s^{\prime}\in\mathcal{S}}\bar{p}(s^{'}|s,\pi_{k}^{*}(s))V_{k}(s^{\prime})\}$; 
\ENDIF 
\ENDFOR
\STATE Let $k=k+1$ and go to step 2 until $\pi_{k}^{*}(s)=\pi_{k+1}^{*}(s)$.% for all $s\in S$.
\STATE \textbf{Return:} The optimal policy $\pi^{*}$.
\end{algorithmic}
\end{algorithm}

\addtolength{\topmargin}{.05in}
\vspace{-0.2em}
\section{Simulation Results} \label{Sec:Section 4}
\vspace{-0.2em}
In this section, we conduct extensive simulations to evaluate the performance the optimal policy. We compare it against two benchmark policies, i.e., the random policy and the greedy policy.  Under the random policy, the decision maker randomly selects actions at each decision epoch. In the greedy policy, the decision maker chooses the action that minimizes the expected post-action TAoI at each time step. The expected post-action TAoI is defined as the expected TAoI after the corresponding action taken in time step $t$. For instance, the expected post-action TAoI of action $(0, 0)$ is given by $(1 - p_s)(\Delta + T_1^u + T_1^c) + p_s(T_1^u + T_1^c)$. The simulation parameters $T_{1}^{u}$, $T_{2}^{u}$, $T_{1}^{c}$, and $T_{2}^{c}$ are set such that $T_{1}^{u}<T_{2}^{u}$ and $T_{1}^{c}<T_{2}^{c}$. The inference accuracy for both SLM and LLM varies between 0.05 and 0.99 \cite{Li2023}.

\begin{figure}[!t]
\centering 
\subfloat[Average TAoI versus $T_{2}^{u}$ ($T_{1}^{u}=1$, $T_{1}^{c}=2$,
$T_{2}^{c}=11$).]{\includegraphics[width=0.38\textwidth]{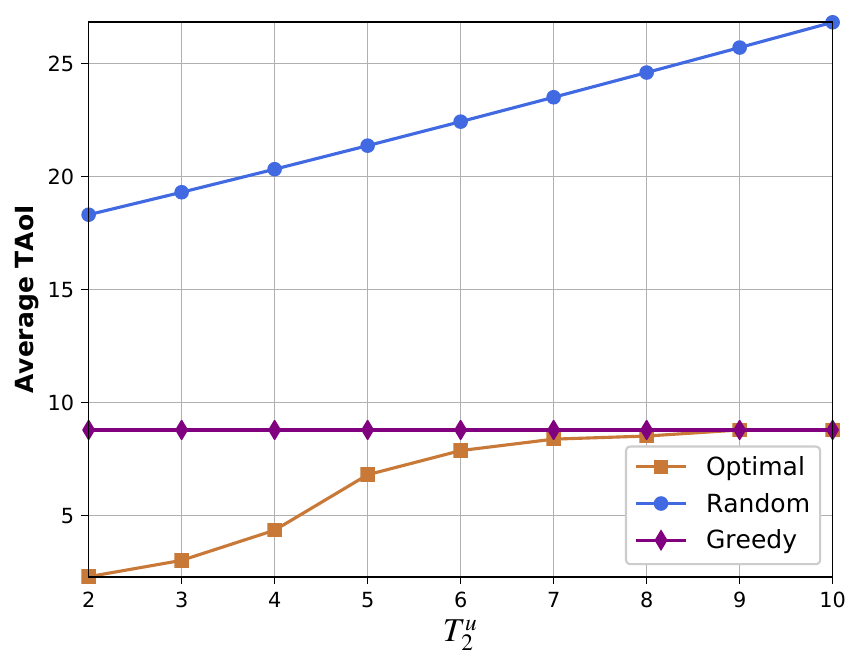}
}\hfill{}
\subfloat[Average TAoI versus $T_{1}^{c}$ ($T_{1}^{u}=1$, $T_{2}^{u}=2$,
$T_{2}^{c}=11$).]{\includegraphics[width=0.38\textwidth]{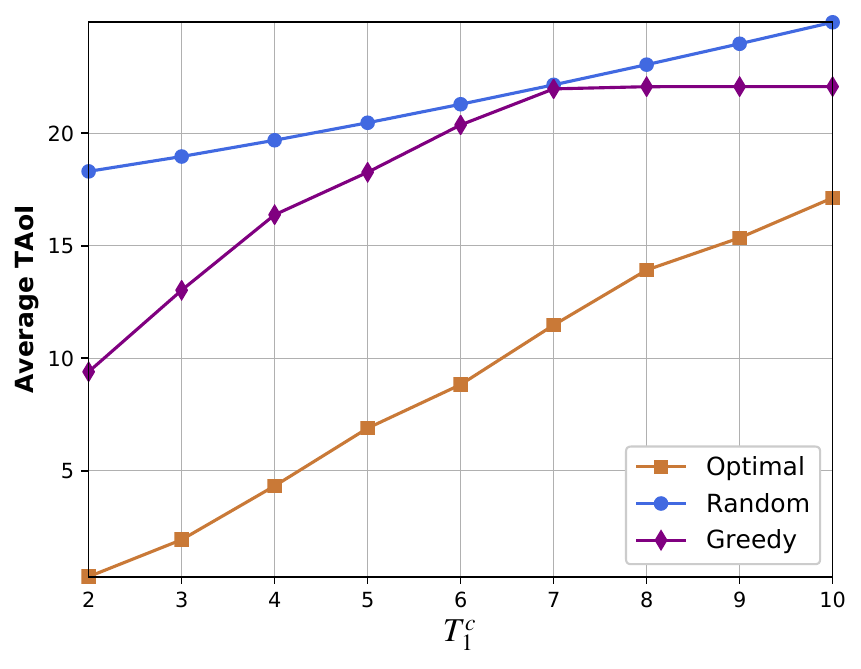}}
\vspace{-0.4em}
\caption{Average TAoI versus $T_{2}^{u}$ or $T_{1}^{c}$ ($p_{s}=0.4$, $q_{s}=0.5$,
$p_{l}=0.6$, $q_{l}=0.8$).}
\label{Fig:Avg_TAoI_Tuc}
\end{figure}

Fig. \ref{Fig:Avg_TAoI_Tuc} compares the average TAoI between the optimal policy and the two baseline policies with respect to the transmission latency $T_{2}^{u}$ and the inference latency $T_{1}^{c}$. 
While the optimal policy's average TAoI generally increases with both latency parameters, its behavior differs markedly between the two cases. For transmission latency $T_{2}^{u}$, shown in Figure \ref{Fig:Avg_TAoI_Tuc}(a), the average TAoI plateaus once $T_{2}^{u}$ exceeds 9, as the optimal policy adaptively avoids high-resolution image selection, making further increases in $T_{2}^{u}$ inconsequential. In contrast, Figure \ref{Fig:Avg_TAoI_Tuc}(b) shows that the average TAoI continues to rise with inference latency $T_{1}^{c}$ without stabilizing, since $T_{1}^{c}$ remains below $T_{2}^{c}$ and thus continues to influence the system's performance through the optimal policy's decision-making process.
%We can see from the figures that the average TAoI of the optimal policy increases with the increasing of transmission latency $T_{2}^{u}$ or inference latency $T_{1}^{c}$. However, as shown in Fig. \ref{Fig:Avg_TAoI_Tuc}(a), the average TAoI of the optimal policy stabilizes after $T_{2}^{u}$ surpasses $9$. This is because, as the transmission latency of high-resolution images $T_{2}^{u}$ increases, the optimal policy tends to avoid choosing high-resolution images, rendering further increments in $T_{2}^{u}$ irrelevant to the average TAoI. The average TAoI of the optimal policy in Fig. \ref{Fig:Avg_TAoI_Tuc}(b) does not level off at the end. This is due to $T_{1}^{c}$ being smaller than $T_{2}^{u}$, meaning $T_{1}^{c}$ will continue to affect the average TAoI of the optimal policy.
Moreover, as shown in Fig.  \ref{Fig:Avg_TAoI_Tuc}(a), when $T_{2}^{u}$ is large, the optimal policy coincides with the greedy policy, which always selects action $(0, 0)$ in this setup. As $T_{2}^{u}$ increases, the optimal policy favors transmitting low-resolution images. Also, given that $T_{1}^{c}$ is substantially lower than $T_{2}^{c}$, the potential benefits of LLM processing become outweighed by its latency costs. These combined effects drive the optimal policy to naturally align with the greedy policy.

%Finally, unlike the consistent stability of the greedy policy in Fig. \ref{Fig:Avg_TAoI_Tuc}(a), Fig. \ref{Fig:Avg_TAoI_Tuc}(b) shows that as $T_{1}^{c}$ increases, the greedy policy gradually stabilizes. This is because the parameter configuration in Fig. \ref{Fig:Avg_TAoI_Tuc}(a) leads the greedy policy to always choose action $(0, 0)$, so the increase in $T_{2}^{u}$ does not affect its choice. However, in Fig. \ref{Fig:Avg_TAoI_Tuc}(b), the parameters lead the greedy policy to choose action $(0, 0)$ or $(1, 0)$ when $T_{1}^{c}$ is small, but it shifts to $(1, 1)$ as $T_{1}^{c}$ increases.

\begin{figure}[!t]
\centering \subfloat[Average TAoI versus $p_{s}$ ($q_{s}=0.5$, $p_{l}=0.6$, $q_{l}=0.8$).]{\includegraphics[width=0.38\textwidth]{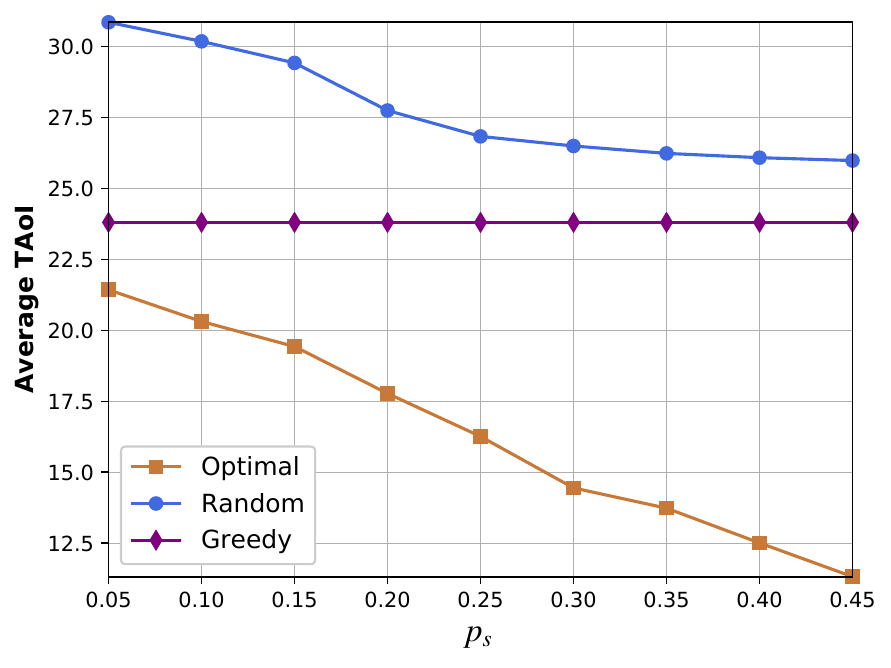}}\hfill{}
\subfloat[Average TAoI versus $q_{l}$ ($p_{s}=0.3$, $q_{s}=0.4$, $p_{l}=0.5$).]{\includegraphics[width=0.38\textwidth]{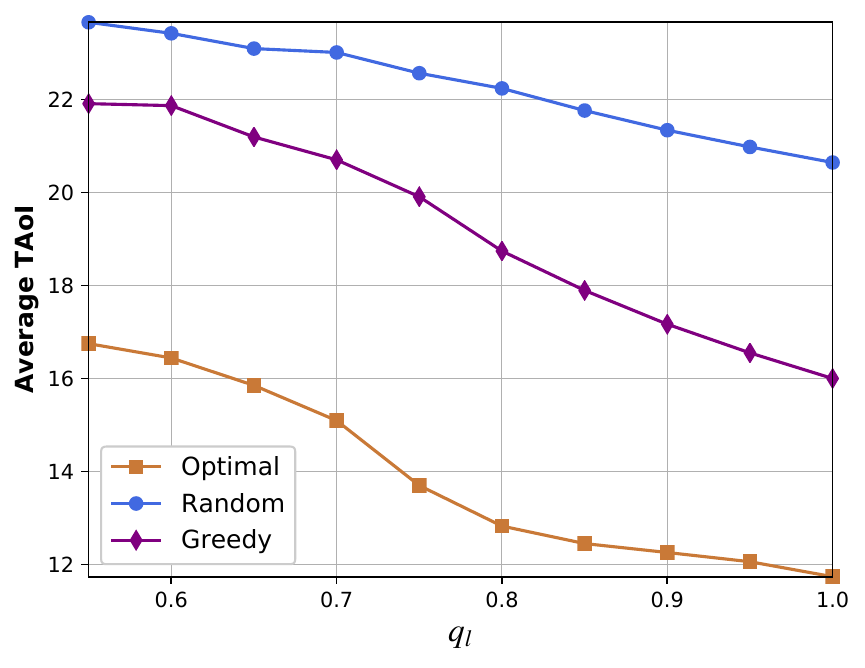}}
\vspace{-0.4em}
\caption{Average TAoI versus $p_{s}$ or $q_{l}$ ($T_{1}^{u}=3$, $T_{2}^{u}=4$,
$T_{1}^{c}=8$, $T_{2}^{c}=10$).}
\label{Fig:Avg_TAoI_qspl}
\end{figure}

In Fig. \ref{Fig:Avg_TAoI_qspl}, the average TAoI of the optimal policy and the two baselines are compared with respect to model accuracy parameters $p_{s}$ or $q_{l}$. As shown in Fig. \ref{Fig:Avg_TAoI_qspl}(a) and Fig. \ref{Fig:Avg_TAoI_qspl}(b), we can see that the optimal policy consistently achieves lower average TAoI compared to baseline policies. Moreover, the average TAoI of the optimal policy decreases with the increase of $p_{s}$ or $q_{l}$, which indicates that, enhanced model accuracy, regardless of resolution or model size, is beneficial for the success of inference tasks. The impact of these parameters, however, manifests differently. Fig. \ref{Fig:Avg_TAoI_qspl}(a) shows that as $p_{s}$ increases, the average TAoI of the optimal policy exhibits a sharp decrease. This pronounced improvement occurs because action $(0,0)$, which offers the lowest latency, becomes increasingly favored by the optimal policy as its inference accuracy improves. In contrast, Figure \ref{Fig:Avg_TAoI_qspl}(b) shows that while increases in $q_l$ also reduce the average TAoI, this reduction occurs more gradually and diminishes at higher values of $q_l$, suggesting a point of diminishing returns in the accuracy-latency trade-off.
%This is because action $(0,0)$ has the lowest latency, and as the inference accuracy of $(0,0)$ increases, the optimal policy tends to favor action $(0,0)$, making its advantage more pronounced.
%As shown in Fig. \ref{Fig:Avg_TAoI_qspl}(b), the average TAoI of the optimal policy exhibits a sharp decrease with $q_{l}$ gradually. When $q_{l}$ is significantly high, the reduction in the average TAoI is not evident.

%when $q_{l}$ significantly exceeds the inference accuracy of other actions, the rate of decline in the average TAoI for the optimal policy initially increases and then slows down. This trend is related to the changes in the frequency of selecting action $(1, 1)$ within the optimal policy. Specifically, the greater the change in the frequency of selecting action $(1, 1)$, the faster the decline in the average TAoI.
%Lastly, Fig. \ref{Fig:Avg_TAoI_qspl}(b), reveals that the rate at which the average TAoI of the optimal policy decreases slows down as  $q_{l}$ significantly surpasses the inference accuracy of other actions. This is because as the accuracy of action $(1,1)$ increases, the optimal policy gradually favors this action. Given that its latency remains constant, the average TAoI declines at a moderated pace.

\section{Conclusions}\label{Sec:Section 5} 
%\vspace{-0.4em}
In this paper, we introduced a novel remote inference system that combines SLM and LLM to optimize both accuracy and timeliness. We developed a dynamic control policy that minimizes the TAoI through joint optimization of resolution and inference decisions. By formulating the control problem as an infinite-horizon SMDP and transforming it into an equivalent MDP, we proved that the optimal control policy follows a threshold structure. Building on this insight, we developed a RPI algorithm that leverages this threshold structure to efficiently determine the optimal policy while minimizing computational overhead. Our extensive simulation results demonstrated the superiority of our proposed approach, with the optimal policy consistently achieving lower average TAoI compared to existing benchmark policies.
%In this paper, we proposed a remote inference system with hybrid SLM and LLM and studied its dynamic control policy to ensure the accuracy and timeliness of inference. We used TAoI as a performance metric and minimized the average TAoI by jointly optimizing the resolution decision and an inference decision. The dynamic control problem was formulated as an infinite-horizon SMDP, which was then transformed into an equivalent MDP with uniform time steps. Then, we proved that the optimal control policy followed a threshold structure and proposed a low-complexity RPI algorithm that leveraged this structure to derive the optimal policy. Simulation results showed that the proposed optimal policy achieved a lower average TAoI compared to benchmark policies.

\appendix{}
\vspace{-0.4em}
\subsection{Proof of Theorem 1} \label{PR:proof_Theorem1} 
First, we define $Q'(\Delta_{2},\Delta_{1}, \bm{a})=Q(\Delta_{2},\bm{a})-Q(\Delta_{1},\bm{a})$ for convenience. For any $\Delta_{1}$, $\Delta_{2}\in\mathcal{S}$ with $\Delta_{1}\leq\Delta_{2}$, we have 
\begin{align}
 & Q'(\Delta_{2},\Delta_{1},\hat{\bm{a}})-(V(\Delta_{2})-V(\Delta_{1}))\nonumber \\
 = &\Delta_{2}-\Delta_{1}-\frac{\epsilon}{L(\hat{\bm{a}})}(V(\Delta_{2})-V(\Delta_{1}))\nonumber \\
 & \quad+\frac{\epsilon(1-p)}{L(\hat{\bm{a}})}(V(\Delta_{2}+L(\hat{\bm{a}}))-V(\Delta_{1}+L(\hat{\bm{a}}))).
\end{align}
Given that the concavity of $V(s)$ is established in Lemma \ref{LM:Lemma2}, it follows that $V(\Delta_{2}+L(\hat{\bm{a}}))-V(\Delta_{1}+L(\hat{\bm{a}}))\leq V(\Delta_{2})-V(\Delta_{1}))$. Then, we can get that 
\begin{align}
 &  Q'(\Delta_{2},\Delta_{1},\bm{\hat{\bm{a}}})-(V(\Delta_{2})-V(\Delta_{1})) \nonumber \\
 &\leq\Delta_{2}-\Delta_{1} +\frac{\epsilon(1-p)}{L(\bm{\hat{\bm{a}}})}(V(\Delta_{2})-V(\Delta_{1}))\nonumber \\
 &\quad-\frac{\epsilon}{L(\bm{\hat{\bm{a}}})}(V(\Delta_{2})-V(\Delta_{1}))\nonumber \\
  &=\Delta_{2}-\Delta_{1}-\frac{\epsilon p}{L(\bm{\hat{\bm{a}}})}(V(\Delta_{2})-V(\Delta_{1})).
\end{align}
As shown in Lemma \ref{LM:Lemma3}, we have $V(\Delta_{2})-V(\Delta_{1})\geq\frac{L(\bm{\hat{\bm{a}}})}{\epsilon p}(\Delta_{2}-\Delta_{1})$. This implies that $Q'(\Delta_{2},\Delta_{1})-(V(\Delta_{2})-V(\Delta_{1}))\leq0$. 

Next, we prove the threshold structure of the optimal policy. Suppose $\Delta_{2}\geq\Delta_{1}$ and $\pi^{*}(\Delta_{1})=\bm{\hat{\bm{a}}}$,
we have $V(\Delta_{1})=Q(\Delta_{1},\bm{\hat{\bm{a}}})$,
i.e., $V(\Delta_{1})=Q(\Delta_{1},\bm{\hat{\bm{a}}})$. It is straightforward to obtain $V(\Delta_{2}) \geq Q(\Delta_{2}, \hat{\bm{a}})$, since $V(\Delta_{2}) - V(\Delta_{1}) \geq Q(\Delta_{2}, \hat{\bm{a}}) - Q(\Delta_{1}, \hat{\bm{a}})$. Moreover, since the value function is a minimum of two state-action value functions, we have $V(\Delta_{2})\leq Q(\Delta_{2},\bm{\bm{\hat{\bm{a}}}})$. Therefore, we can conclude that $V(\Delta_{2}) = Q(\Delta_{2}, \hat{\bm{a}})$ and that $\pi^{*}(\Delta_{2}) = \hat{\bm{a}}$. This concludes the proof.

\bibliographystyle{IEEEtran}
\bibliography{IEEEabrv,CoT_Ref}

\end{document}